\documentclass[aps,pre,preprint,superscriptaddress, showpacs]{revtex4}
\usepackage{times}
\usepackage{graphicx}
\usepackage{psfrag}
\usepackage{ae}
\usepackage{amsmath,amssymb}

\begin{document}
\date{\today}

\author{J. R. Bordin} 
\affiliation{Instituto de F\'{\i}sica, Universidade Federal
do Rio Grande do Sul\\ Caixa Postal 15051, CEP 91501-970, Porto Alegre, RS, Brazil}
\author{A. Diehl}
\affiliation{Departamento de F\'{\i}sica, Instituto de F\'{\i}sica e Matem\'atica,
Universidade Federal de Pelotas, Caixa Postal 354, CEP 96010-900, Pelotas, RS, Brazil}
\author{M. C. Barbosa} 
\affiliation{Instituto de F\'{\i}sica, Universidade Federal
do Rio Grande do Sul\\ Caixa Postal 15051, CEP 91501-970, 
Porto Alegre, RS, Brazil}
\author{Y. Levin} 
\affiliation{Instituto de F\'{\i}sica, Universidade Federal
do Rio Grande do Sul\\ Caixa Postal 15051, CEP 91501-970, 
Porto Alegre, RS, Brazil}

\title{Ion fluxes through nano-pores and transmembrane channels}

\begin{abstract}

We introduce an implicit solvent Molecular Dynamics approach for calculating 
ionic fluxes through narrow nano-pores and transmembrane channels.
The method relies on a dual-control-volume grand-canonical molecular dynamics (DCV-GCMD) simulation 
and the analytical  solution for the electrostatic potential inside a cylindrical nano-pore recently 
obtained by Levin~[Europhys. Lett., {\bf 76}, 163 (2006)]. The theory is used to calculate the ionic fluxes through an artificial 
transmembrane channel which mimics the antibacterial gramicidin A channel.  Both
current-voltage and current-concentration relations are calculated under various experimental 
conditions. We show that our results are comparable to the characteristics associated to the gramicidin A pore, 
specially the existence of two binding sites inside the pore and the observed saturation in the 
current-concentration profiles.

\end{abstract}


\maketitle

\section{Introduction}
\label{Introd}

Ion channels are structures formed when specific proteins are incorporated into the phospholipid 
membrane~\cite{Hille}. The channels serve to establish an electrostatic potential gradient across the
cell membrane by allowing an ion specific flux to pass through the membrane. There are many different ion
channels in living cells.  They differ in composition, pore 
structure, and ion selectivity~\cite{Doyle98,Roux01}. Thus, a full description of the 
architecture and operation of ion channels is a very difficult task. 
From purely electrostatic point of view, operation of ion channels presents an interesting theoretical
puzzle. Since the channel passes through a low-dielectric membrane, there exists a large potential energy
barrier for ionic solvation inside the pore. Yet, in practice, it is well known that when open, ion channels sustain 
a very large ionic transport rate, compatible with a free diffusion.

A recent advance in nanoscale technology is the synthesizing of nanotubes with chemical modified surface~\cite{Harris99}, 
which allows to construct functionalized nanotubes that mimic the behavior of biological ion
channels and are stables when inserted in a lipid bilayer~\cite{Shi08, Hilder10}, and also exhibits
a high energy barrier~\cite{Beu10}. An excellent review of recent theoretical advances in synthetic nanotubes 
which mimic the behavior of biological ion channels is given by Hilder~{\it{et al}}~\cite{Hilder_rev11}. 

There have been a number of different strategies proposed to understand the ion transport across biological
and synthetic channels. One approach used extensively over the last few years is all-atom molecular dynamics 
simulation (MD). The advantage of atomistic MD is that molecular structure of the pore, ionic species, and water are 
taken explicitly into account~\cite{Roux01, Beu10, Hilder_rev11,Roux96,Edwards02,Rabi04, Joseph03, Majunder07, Hilder11B}. 
Although this method is very appealing, it remains a huge task to relate the measured observables to the 
experimental results. One of the main obstacles are the computational times needed to achieve the time scales 
observed experimentally~\cite{Levitt99,Roux02,AKR2005,Aksime05} which, in most cases, requires large-scale MD 
simulations in special-purpose machines~\cite{anton2010}. Another problem is the choice of the molecular 
force field to be used. For instance, in the case of the biological channel gramicidin A (gA), 
one of the most studied ionic channels, the first MD simulations 
have predicted a potential of mean-force (PMF) with a barrier much larger than expected experimentally~\cite{AKR2005}. 
Recent MD simulations with molecular force field properly constructed appear 
to produce PMFs in semiquantitative agreement with experiments~\cite{Allen061,Allen062,Ing11}. 
Although promising, this agreement seems to be very sensitive to the molecular force field used and with 
the membrane where the gA channel is embedded~\cite{Ing11}. 
Furthermore, the classical water models used in all-atom simulations are 
parametrized for bulk water and might show erroneous behavior in a strongly confined environment. Such  
artifacts of classical water and ion models have been recently observed in the studies of ionic solvation 
near interfaces, when compared to the full {\it ab initio} simulations~\cite{Baer2011}. 
It has been shown that for ionic solvation in an interfacial geometry properly constructed dielectric 
continuum models agree better with results of full {\it ab initio} simulations than the classical 
explicit water models~\cite{Levin2009,dosSantos2010,Baer2011}. 

One alternative to atomistic MDs are the, so called, Brownian dynamics 
simulation (BD)~\cite{Moy00,Corry00,Edwards02}. In these simulations only ionic movement is integrated, 
while the protein degrees of freedom are held fixed and the water is treated as a 
uniform dielectric continuum. This significantly reduces the computational cost of the simulation, allowing to 
access much larger time scale. Nevertheless, BD simulations still 
requires solution of Poisson partial-differential equation at each new time step of the simulation, 
making them quite difficult to implement.  
Recently new atomic-resolution BD simulations have been proposed~\cite{Carr11}. Using PMFs derived from 
all-atom MD simulations, this method was able to access large simulation times at reasonably low computational cost~\cite{Carr11}.

Thus, in order to avoid the difficulties of all-atom and 
BD simulations, as well as to be able to explore large time scales necessary for measuring the transmembrane currents, 
it might be useful to use the dielectric continuum models of 
ion channels as a first order approximation for the transmembrane dynamics.  
In this paper we  explore the transmembrane ion fluxes using the 
continuum electrostatics model recently introduced by Levin~\cite{Levin2006}. Briefly stated, Levin
solved the Poisson equation with the appropriate boundary conditions 
to obtain an analytical expression for the electrostatic interaction potential (Green function) 
between the charges inside a finite cylindrical pore passing through a low-dielectric phosphoric membrane.    
This electrostatic potential can be used to calculate the forces acting between the ions inside the pore
and between the ions and the charged protein residues embedded in the cell membrane. 
It is our goal in this study to use the interaction potential derived by Levin in a series of Dual-Control-Volume 
Grand-Canonical Molecular Dynamics (DCV-GCMD)~\cite{DCVGCMD94,DCVGCMD98} simulations of a 
simple model of narrow transmembrane channel.
We should stress that our model is very different from the mean-field Poisson-Nernst-Plank (PNP) 
theory~\cite{Eisenberg99,Luchinsky09}, where the Poisson equation and the continuity equations for 
mobile ions are solved simultaneously in a self-consistent way.  For narrow channel such as gA the correlation
effects between the ions are of fundamental importance and the mean-field description of ionic conduction is bound to
fail~\cite{Levin02}. 

Our simulations are performed for different electrolyte concentrations and external applied electric fields
to verify if our model reproduces the ionic flow obtained  for one specific system: the gramicidin A channel. 
The comparision between our model and experiments for various concentrations and external fields is 
an indication that it is able to reproduce the observed ionic fluxes in narrow channels~\cite{Hilder_rev11, Hilder11}. 

The paper is organized as follows. The model and computational details are given in Sec.~\ref{sec1}. 
The results are discussed in Sec.~\ref{discussion} and the summary and conclusions are presented in 
Sec.~\ref{conclusion}. 

\section{The model system and the simulation methodology}
\label{sec1}
 \subsection{A model for transmembrane channels}

\begin{figure}[t]
\begin{center}
\includegraphics[width=12cm]{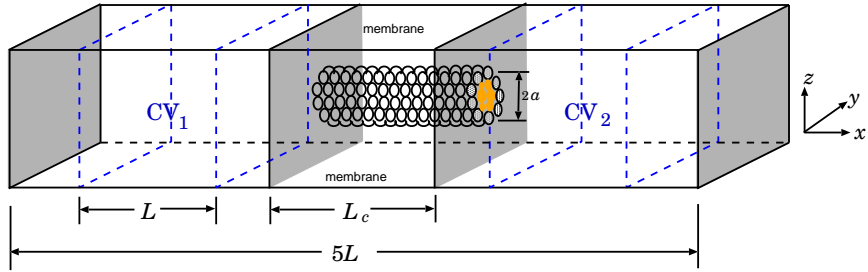}
\end{center}
\caption{Schematic depiction of the simulation cell. The ionic channel is built as a cylinder made of a 
sequence of annular rings of LJ spheres. The system is confined in $x$ direction by two flat walls (represented by gray) 
at each end of the simulation box. The same flat walls are used to maintain the flux of cations between the two reservoirs (CVs) 
through the channel. Periodic boundary conditions are applied in both $y$ and $z$ directions.}
\label{fig1}
\end{figure}

We use the channel-reservoir setup illustrated schematically in Fig.~\ref{fig1} 
to calculate the ionic currents due to electrostatic potential gradients. 
The simulation box, a cubic parallelepiped with volume $5L\times L\times L$ and $L = 20\,$ {\AA}, contains 
two reservoirs (two control volumes CV$_1$ and CV$_2$, within which the chemical potentials 
of the ionic species are maintained constant) and a membrane with a channel connecting these two 
reservoirs through small buffers regions. The channel structure, a simplified model of a transmembrane pore, 
is built as a cylindrical tube, with radius $a=3\,\mbox{ \AA}$ and length 
$L_c=35\,$ \AA, made of stationary Lennard-Jones (LJ) spheres of diameter $\sigma_c=2\,$ \AA.

Both sides of the channel structure --- except for the orifices --- are bound by confining walls, 
as well as the two extremes of the simulation box in the $x$ direction, see Fig.~\ref{fig1}. 
The system contains positive and negative ions with 
diameters $\sigma_+$ and $\sigma_-$, respectively,  inside a structureless solvent of 
dielectric constant $\epsilon_w = 80$ --- the same value of dielectric is used inside and outside the channel ---
while the membrane has a dielectric constant equals to 
$\epsilon_m = 2$, both in units of vacuum permittivity. 
It should be noted that in confined environments the dielectric constant of water could be considerably 
lower than the bulk value 80~\cite{Allen2004,Li2010}. 
However, in this work we follow the same prescription of previous dielectric continuum models, where the use of 
same dielectric constant is compensated by fixing a lower diffusion constant for the positive ions moving inside the 
channel~\cite{Edwards02,Kurnikova99}.

The particle-particle interactions are separated into short and long-range contributions, while the 
particle-channel has only a short-range interaction. 
For the short-range part we will use the WCA LJ potential~\cite{AllenTild} 
\begin{equation}
\label{LJCS}
U_{ij}^{\rm{WCA}}(r) = \left\{ \begin{array}{ll}
U_{{\rm {LJ}}}(r) - U_{{\rm{LJ}}}(r_c)\;, \qquad r \le r_c\;, \\
0\;, \qquad \qquad \qquad \qquad \quad r  > r_c\;,
\end{array} \right.
\end{equation}
where $U_{\rm LJ}(r)$ is the standard 12-6 LJ potential. The cut-off distance is $r_c = 2^{1/6}\sigma_{ij}$, 
where $\sigma_{ij} = (\sigma_i + \sigma_j)/2$ is the center-to-center 
separation between an ion of species $i$ (cation or anion) and a particle of species $j$ 
(cation, anion, or a fixed LJ channel sphere) separated by a distance $r$. 
The confining walls in simulation box extremes and surrounding the channel structure
are modeled with the same WCA LJ potential, however considering the $x$-projection of 
the distance between one ion in the bulk and the wall position. 
The long-range contribution is calculated depending on the region where the ion is located. 
For the regions outside the channel, the interaction energy between the two ions is the usual Coulomb potential
\begin{equation}
\label{coulomb}
U_{ij}^{\rm coul}(r) = \frac{1}{4\pi \epsilon_w}\,\frac{q_i\,q_j}{r_{ij}}\;,
\end{equation}
where $r_{ij}$ is the distance between the two ions.  The infinite extent of the particle reservoir is
taken into account using the Ewald summation. 

For the region inside the channel we will use the model introduced by Levin~\cite{Levin2006}.  In this model ions
inside the pore interact through the electrostatic potential which is only a function of their separation in the 
$x$-direction (along the axis of symmetry of the channel).  This is quite reasonable for narrow channels that
allows only single file motion. 
When one ion with charge $q$ enters into the channel it interacts with the other ions and with the $p$ residues of charge $-q$ 
embedded into the channel wall at transverse distance $\rho$ from the central axis. In addition to the electrostatic interaction 
between the ions and the residues, there is a self-energy penalty $U$ associated with an ion entering into the 
region surrounded by the low-dielectric material.  The electrostatic energy of the ions inside the pore is then
\begin{equation}
\label{totalenergy}
V=\frac{1}{2}\sum_{i=1}^N\sum_{j=1}^Nq_i\varphi_{\rm in}(x_i,\,x_j)+\sum_{i=1}^N\sum_{j=1}^p q_i\varphi_{\rm out}(x_i,\rho,x_j)+
\sum_{i=1}^Nq_i\,U(x_i)\;,
\end{equation}
where $\varphi_{\rm in}(x_i,\,x_j) = \varphi_1(x_i,\,x_j) + \varphi_2(x_i,\,x_j)$. The two electrostatic potentials are~\cite{Levin2006}
\begin{eqnarray}
\label{phi1}
&&\varphi_1(x_i,\,x_j) = \nonumber\\
&&\frac{q}{\epsilon_w}\int_0^{\infty} dk \frac{\{\alpha^2(k)e^{k|x_i-x_j|-2kL_C} + \alpha(k)\beta(k)[e^{-k(x_i+x_j)}
+e^{k(x_i+x_j)-2kL_c)}]+\beta^2(k)e^{-k|x_i-x_j|} \}} {\beta^2(k) - \alpha^2(k)\exp [-2kL_C]}\;,\nonumber\\
\end{eqnarray}
where $\alpha = [k - (k^2 + \kappa^2)^{1/2}]$ and $\beta =  [k + (k^2 + \kappa^2)^{1/2}]$, with $\kappa$ the 
inverse Debye length that characterizes the electrolyte concentration in the two reservoirs, and 
\begin{equation}
\label{phi2}
 \varphi_2(x_i,x_j) = \frac{4q(\epsilon_w-\epsilon_p)}{\epsilon_w L_c} \sum_{n=1}^{\infty} \frac{K_0(k_na)K_1(k_na)\sin(k_nx_i)\sin(k_nx_j)}
 {\epsilon_w I_1(k_na)K_0(k_na)+\epsilon_p I_0(k_na)K_1(k_na)}\;,
\end{equation}
where $I_n$ and $K_n$ are the modified Bessel functions of order $n$.
Here $a$ is the channel radius and $k_n = n\pi/L_c$. The interaction potential between an ion and a fixed charged residue is
\begin{equation}
\label{phiout}
 \varphi_{\rm out}(x_i,\rho,x_j) = \frac{4q}{L_c} \sum_{n=1}^{\infty}\frac{1}{k_na} \frac{K_0(k_n\rho)\sin(k_nx_i)\sin(k_nx_j)} 
{\epsilon_w I_1(k_na)K_0(k_na)+\epsilon_p I_0(k_na)K_1(k_na)}\;.
\end{equation}
Finally, the electrostatic potential responsible for the barrier is given by
\begin{eqnarray}
\label{barrier}
U(x)&=&\frac{q}{2\epsilon_w}\int_0^{\infty} dk \frac{\{2\alpha^2(k)e^{-2kL_C} + \alpha(k)\beta(k)[e^{-2kx}+e^{2k(x-L_c)}] \}}
 {\beta^2(k) - \alpha^2(k)\exp [-2kL_C]}+\nonumber\\
&& +\frac{2q(\epsilon_w-\epsilon_p)}{\epsilon_w L_c} \sum_{n=1}^{\infty} \frac{K_0(k_na)K_1(k_na)\sin^2(k_nx)}
 {\epsilon_w I_1(k_na)K_0(k_na)+\epsilon_p I_0(k_na)K_1(k_na)}+\frac{q\kappa}{2\epsilon_w}\;.
\end{eqnarray}

In addition we will place two residues of charge $-q$ each embedded into the channel wall ($\rho=3$ \AA) 
at positions $x = -10.5\,${\AA} and $x = 10.5\,${\AA}, to represent the two binding sites of 
our transmembrane channel. The residues are placed inside the 
channel wall leaving space for the mobile ions to flow.
The channel length and radius are comparable with the size of nanotubes 
and other models for the biological channels such as gramicidin A (gA)~\cite{Edwards02, Hilder11}.
Although the gA channel does not have explicit charged residues,
but only dipolar carbonyl groups~\cite{Ketchem97, AKR2005}, these behave similarly to the 
charged binding sites used in our model~\cite{Edwards02}. Furthermore, recent studies have 
shown that it is possible to use functionalized synthetic nanotubes to 
obtain ionic fluxes similar to the ones observed in gA channel~\cite{Hilder11}. 

\subsection{Simulation details}

Ion channels can be highly selective to the cation passage, and nanotubes can be functionalized
to mimic this behavior~\cite{Hille, Hilder_rev11}. Although the actual transport process is very complicated, we can identify three main 
steps~\cite{AKR2005}: (1) cation entry, where the positive ions are dehydrated, (2) cation translocation 
through the channel, and (3) cation exit. Since in our model we do not consider water molecules explicitly, we simulate step (1) above 
using a diameter for the cations equals to $\sigma_+ = 2\,$\AA, while the negative ions have a diameter 
$\sigma_- = 4\,$\AA. Since the channel is very narrow, the available radius for ion movement is approximately $a-\sigma_c/2$ in Fig.~\ref{fig1}. 
Therefore, using these diameters only positive ions 
can enter into the channel. In all interactions the energy parameter of
LJ potential was defined as $\epsilon = 1 k_BT$.
During the MD steps, we have used Langevin dynamics to simulate the effect of solvent on the cation 
and anion movement, solving the equation of motion of ion $i$,
\begin{equation}
\label{ResFor}
m_i\,\vec{a}_i = \vec{F}_i - m\gamma \vec{v}_i + \vec{W}(t)\;,
\end{equation}
where $\vec{F}_i$ is the total force on ion $i$ due to all entities explicitly present in the 
model (other ions, protein residues, and walls), $\gamma$ is the friction
coefficient, and $\vec{W}(t)$ is the random force~\cite{Wiener1923} due to solvent. The temperature of the system is maintained 
constant using the fluctuation-dissipation theorem,
\begin{equation}
\langle \vec{W}(t).\vec{W}(t')\rangle = 6k_BT\gamma \delta (t-t')\;,
\end{equation}
which relates the friction coefficient to the fluctuations of the random force using the 
appropriate $\gamma$ value. In our simulations we have used $m_i=6.5\times 10^{-26}$kg, corresponding to the mass of $K^+$ ion;
$\gamma^{-1} = 53\,$fs for the region outside the channel, 
corresponding to diffusion constant $D=k_B T/m_i \gamma = 3.37\times 10^{-9}\,$m$^2$/s; 
and $\gamma_c^{-1} = \gamma^{-1}/3$ inside the channel corresponding to 
diffusion constant of  $D=1.12\times 10^{-9}\,$m$^2$/s. 
These values were chosen in order to maintain the temperature fixed at 298 K and to reproduce the experimental behavior, 
particularly the saturation observed in the current-concentration curves~\cite{Edwards02}.

We apply a linear voltage gradient across the membrane from the right border of CV$_1$ to the left border of CV$_2$. 
The concentrations in both reservoirs are maintained constant using DCV-GCMD 
simulations~\cite{DCVGCMD94,DCVGCMD98,India07,LADERAI,LADERAII,Pohl,Horsh2009}. Briefly stated, in the  DCV-GCMD simulation two 
control volumes (CVs) are initially prepared at desired concentrations using the grand canonical Monte Carlo (GCMC) simulation and 
then evolved in time using the molecular dynamics (MD) simulation. 
Since the dynamics alters the CV concentrations, the MD steps are 
intercalated with the grand canonical Monte Carlo (GCMC) simulations performed inside the two control volumes (CV)
shown in Fig.~\ref{fig1}. This restores the concentrations to their initial values. In our simulations we 
have used 50 GCMC steps for every 500 MD steps. It should be noted that our DCV-GCMD method is closely related to the 
GCMC/BD algorithm proposed by Im {\it et al.}~\cite{Im2000} to simulate the ionic conductance in membrane channels. 

Since our simulation setup is periodic only in $y$ and $z$ directions, as shown in Fig.~\ref{fig1},  for the long-range interactions 
described by Eq.~(\ref{coulomb}) we have used the Ewald summation 
with the implementation of Yeh and Berkowitz~\cite{Berkowitz99} for the slab geometry. 
The equations of motion were integrated using the velocity Verlet algorithm, 
with a time step of 8.0 fs in the MD simulations. The observables were obtained using 5 to 10 independent realizations.

\section{Results and discussion}
\label{discussion}
\begin{figure}[t]
\begin{center}
\includegraphics[width=8cm]{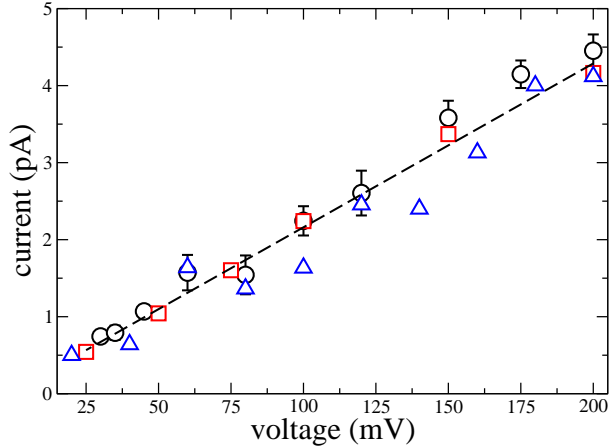}
\end{center}
\caption{Current-voltage curve for 0.5 M concentration of electrolyte in both CVs.
Open circles are our simulations results . The open squares and the 
open triangle are the experimental and BD results for gA channel, respectively, extracted from Fig.~12 of Ref.~\cite{Edwards02}. }
\label{fig2}
\end{figure}

We start our discussion with the two CVs having the same concentration, and the ionic diffusion through the channel driven 
by the externally imposed electrostatic potential gradient between the two CVs. 
We are particularly interested in the current-voltage and current-concentration curves and their 
comparison with the experimental results, mainly the appearance of the experimentally observed saturation in the current-concentration profiles.
In Fig.~\ref{fig2} we show the current-voltage curve for the concentration of 0.5 M in both CVs. As one can see, we obtain the same 
expected linear dependence between the ionic current and applied voltage, observed in experiments and earlier BD simulations 
for gA channel~\cite{Edwards02}. 

\begin{figure}[t]
\begin{center}
\includegraphics[width=8cm]{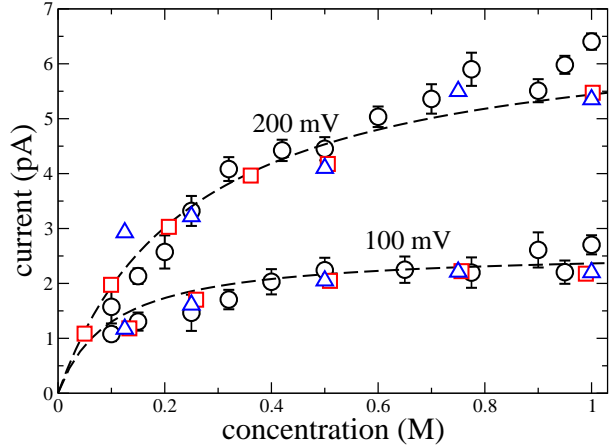}
\end{center}
\caption{Current-concentration curves for external voltages of 100 mV and 200 mV. Our simulations are represented by open circles. 
The open squares and open triangles are the experimental and BD results, respectively, extracted from Fig.~12 of Ref.~\cite{Edwards02}.}
\label{fig3}
\end{figure}

Next we analyze the behavior of the current-concentration profiles for two externally applied voltages of 100 mV and 200 mV,  
Fig.~\ref{fig3}. Compared to the experimental and BD simulation results~\cite{Edwards02}, our model is able to capture the 
same saturation of the ionic current. For 200 mV and above 0.9 M concentration, 
however, we find some deviation from the experimental results. The deviation appears to be an artifact of including only 30 
terms in the infinite series of the electrostatic potential inside the cell, Eqs. (\ref{phi2}) and (\ref{phiout}).  Cutting the 
infinite series at only 30 terms softens the repulsion between the ions inside the channel favoring an increased flux.
 
\begin{figure}[ht]
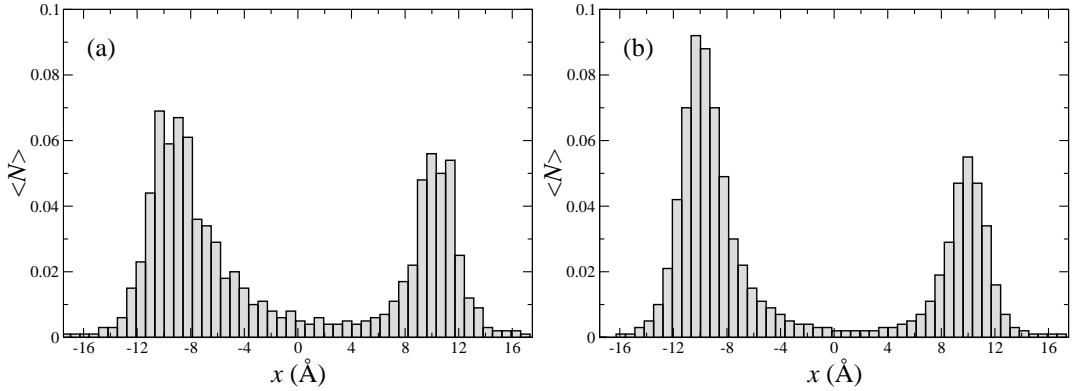

\begin{center}
\includegraphics[width=7cm]{fig4a}
\includegraphics[width=7cm]{fig4b}
\end{center}
\caption{Mean number of cations in the axial direction of gA channel with (a) no applied voltage and (b) 200 mV applied potential. 
In both cases there is 0.5 M monovalent electrolyte solution in the two CVs.}
\label{fig4}
\end{figure}

Experimental~\cite{Tian1999,Urry89,Olah91,Scha78} and BD simulations~\cite{Edwards02} data for gA channel
and theoretical results for functionalized nanotubes~\cite{Hilder11} have proposed two large 
concentration peaks at the binding sites separated by a cation depleted region. This is exactly what we observe  in our 
simulations, as shown in Fig.~\ref{fig4} for 0.5 M monovalent solution in both CVs with no voltage and 200 mV applied 
potential difference. 

\begin{figure}[ht]
\begin{center}
\includegraphics[width=8cm]{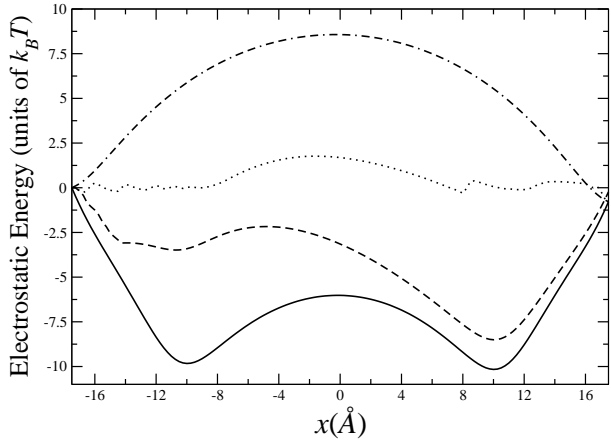}
\end{center}
\caption{Electrostatic energy profile inside the channel obtained from Levin's model~\cite{Levin2006}. 
The curves show the potential of mean force (PMF) felt by an ion moving through a gA-like channel with two charged residues 
at $x=-10.5\,${\AA} and $x=+10.5\,${\AA} (solid line). The PMF inside a  channel which already contains a free cation near 
the first binding site is represented by a dashed line, while the PMF when there are two free cations near the two 
binding sites is represented by a doted line. The dot-dashed line represents the electrostatic potential produced by Eq.~(\ref{barrier}). 
The concentrations in the two CVs are 0.5 M, with no applied voltage.}
\label{fig5}
\end{figure}

To understand better the mechanism of ionic translocation,
in Fig.~\ref{fig5} we plot the electrostatic potential inside the transmembrane channel, 
Eq.~(\ref{totalenergy}), at zero applied voltage and for different occupancy inside the channel and 
residues embedded in the channel structure. These profiles were obtained using a dynamical PMF, where
an ion, called ion*, is pulled through the channel, moving a displacement $\delta x$ each time.
If there are other ions inside the channel (called ions 1, 2, 3, etc) we move the ion*, and then
we allow the system to relax for $4$ ps. During this relaxation time only ions 1, 2, etc. can move, while the
ion* remains fixed. Then we evaluate the mean force on ion*, and perform the next displacement $\delta x$.
The procedure is repeated until ion* leaves the channel. Note that during this process the 
other ions inside the channel will feel the electrostatic repulsion from ion* 
(recall that only cations are allowed inside the channel), and will also be forced out of the channel.

If the channel has no fixed residues and is empty, equation (\ref{barrier}) 
leads to a large electrostatic potential energy barrier of approximately 
$8\,k_BT$, which prevents any cation entrance. On the other hand, if the protein has charged residues embedded 
into the surface of the channel, the scenario changes completely. 
As one can see in solid line in Fig.~\ref{fig5}, with no other cation inside the channel, the ion* feels an energy 
barrier of approximately $4\,k_BT$ separating two deep wells at the positions of the binding sites. 

Entrance of a cation alters drastically the potential energy landscape.  
Suppose that one ion, ion 1, is already inside the channel.  We are interested in the potential 
of mean-force (PMF) that a second ion, ion*, will feel as it moves through the channel. 
This PMF is plotted as a dashed curve in Fig.~\ref{fig5}.  A short time after the
ion 1 has entered  the channel its most probable location is at the first binding site. 
Thus, when ion* enters the
channel, it sees the field of attractive residue partially screened by the ion 1, so that the depth of the potential
well produced by the first residue is significantly smaller.  As the ion* moves farther into the channel, 
it forces the ion 1 to move to the position of the second residue and, eventually, to completely leave the channel.  
Consider now that there are two ions, ion 1 and ion 2, inside the channel, with ion 1 located at the first
binding site and ion 2 at the second binding site. The ion* will then encounter a flat energy landscape
 shown in Fig.~\ref{fig5} by the dotted curve.  This then explains the fast transport of ions through 
the ion channel observed experimentally --- the cations  
present at the binding sites screen both the electric field produced by the residues and by the induced charge on the
channel wall.  In the BD studies~\cite{Edwards02} the saturation was observed as the result
of an imposed double well potential of depth of  $8\,k_BT$ 
and a barrier between the wells of $5\,k_BT$. In the present study, the electrostatic energy landscape was obtained
self-consistently, assuming the existence of two monovalen binding sites inside the channel. 

\begin{figure}[ht]
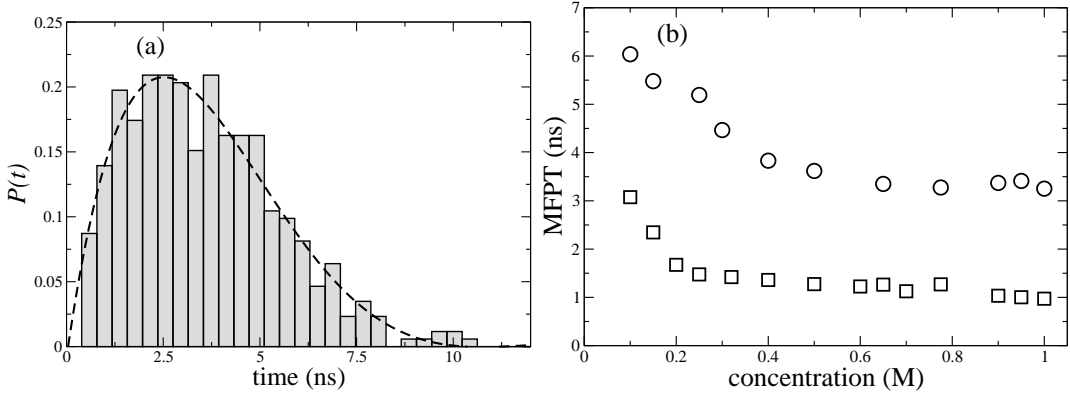

\begin{center}
\includegraphics[width=7cm]{fig6a.eps}
\includegraphics[width=7cm]{fig6b.eps}
\end{center}
\caption{(a) Distribution of passage time $P(t)$ for 100 mV applied voltage and CV concentration of 0.5 M. 
(b) Mean first passage time (MFPT) as a function of CV concentrations for 100 mV (open circles) and 200 mV (open squares) applied voltages.}
\label{fig6}
\end{figure}

In Fig.~\ref{fig6} we show the distribution of times of cation translocation through the channel for voltage difference of
100 mV and CV concentration of 0.5 M. The mean first passage time (MFPT) as a function of 
applied voltage and CVs concentration is also shown in Fig.~\ref{fig6}.  
The saturation observed in Fig.~\ref{fig3} can also be seen in the MFPT results.

\section{Conclusions}
\label{conclusion}

We have used a Dual-Control-Volume Grand-Canonical Molecular Dynamics (DCV-GCMD) simulations to 
study the flow of ions through narrow pores across low dielectric membranes. To account for the electrostatic
interactions inside the channel we have used the analytical potential recently derived by Levin~\cite{Levin2006}. 
For electrolyte concentrations up to 1 M our model is able to reproduce most of experimental behavior of the 
ionic current as a function of electrostatic potential difference. To obtain reliable results
for larger concentrations one must include more terms in the infinite series for the ion-ion interaction potential.
For physiological concentrations of electrolyte, on the other hand, the continuum model presented
in this work appears to show excellent results. The utility of the model is that
the simulations do not require expensive computational resources and can be run on a desktop PC. One interesting
application of the model is to study the dependence of ionic current on mutations of charged residues ---
the position and the charge of the residues that enter into the electrostatic potential can be optimized 
to obtain the desirable channel characteristics.  This could be of interest for
design and implementation of molecular engineered of ion channels and nano-pores, biological or synthetic.    

\section{Acknowledgements}
This work was partially supported by the CNPq, Capes, Fapergs, INCT-FCx, and by the US-AFOSR under the grant FA9550-09-1-0283.

\end{document}